\documentclass[a4paper,accepted=2026-10-14]{quantumarticle}
\pdfoutput=1

\usepackage{graphicx}

\usepackage[utf8]{inputenc}
\usepackage[english]{babel}
\usepackage[T1]{fontenc}	

\usepackage{amsmath}
\usepackage{amssymb}

\usepackage{hyperref}
\hypersetup{
 pdftitle={The three kinds of three-qubit entanglement},
 pdfauthor={Szilárd Szalay},
 pdfsubject={research article on multipartite entanglement of three-qubit states},
 pdfcreator={pdflatex}, pdfproducer={vim},
 pdfkeywords={quantum entanglement} {three-qubit states} {pure states} {entanglement measures},
 unicode=true, pdftoolbar=true, pdfmenubar=true, pdffitwindow=false, pdfstartview={FitH},
 pdfnewwindow=true, colorlinks=true, linktoc=page, linkcolor=blue, citecolor=blue, filecolor=blue, urlcolor=blue
}

\usepackage[numbers,sort&compress]{natbib}

\newcommand{\field}[1]{\mathbb{#1}}

\newcommand{\ket}[1]{\vert #1 \rangle}
\newcommand{\bra}[1]{\langle #1 \vert}

\newcommand{\innerprod}[2]{\langle #1 | #2 \rangle}
\newcommand{\proj}[1]{\ket{#1}\bra{#1}}
\newcommand{\ketbra}[2]{\ket{#1}\bra{#2}}
\newcommand{\Id}{\mathbb{I}}

\newcommand{\LieGrp}[1]{\mathrm{#1}}

\DeclareMathOperator{\tr}{tr}
\DeclareMathOperator{\Det}{Det}
\DeclareMathOperator{\Lin}{Lin}

\newcommand{\set}[1]{\{ #1 \}}

\providecommand{\abs}[1]{{\lvert#1\rvert}}

\providecommand{\norm}[1]{{\lVert#1\rVert}}
\providecommand{\bignorm}[1]{{\big\lVert#1\big\rVert}}

\newcommand{\indexud}[2]{^{#1}_{\phantom{#1}#2}}

\newcommand{\equals}{{\;\;=\;\;}}

\newcommand{\equalsref}[1]{\overset{\eqref{#1}}{\equals}}

\begin{document}
\title{The three kinds of three-qubit entanglement}
\author{Szilárd Szalay}
\email{szalay.szilard@wigner.hu}
\affiliation{Department of Theoretical Solid State Physics,
HUN-REN Wigner Research Centre for Physics, Budapest, Hungary}
\affiliation{Department of Theoretical Physics,
University of the Basque Country UPV/EHU, Bilbao, Spain}
\affiliation{EHU Quantum Center,
University of the Basque Country UPV/EHU, Leioa, Biscay, Spain}

\date{October 16, 2026}

\begin{abstract}
We construct an important missing piece in the entanglement theory of pure three-qubit states,
which is a polynomial measure of W-entanglement.
This works in parallel to the three-tangle,
which is a polynomial measure of GHZ-entanglement,
and to the bipartite concurrence,
which is a polynomial measure of bipartite entanglement.
We also show that these entanglement measures are ordered,
the bipartite measure is larger than the W measure, which is larger than the GHZ measure.
It is meaningful then to consider these three types of three-qubit entanglement, which are also ordered,
bipartite is weaker than W, which is weaker than GHZ,
in parallel to the order of the three kinds of equivalence classes of entangled three-qubit states.
\end{abstract}

\maketitle{}

\section{Introduction}
\label{sec:Intro}

\begin{figure}\centering
\includegraphics{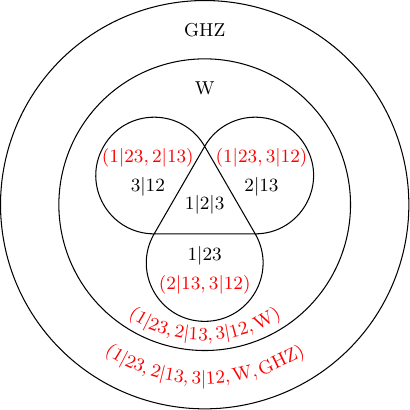}
\caption{SLOCC classification of three-qubit state vectors.
The types of entanglement contained in the classes are written in red in parentheses.}
\label{fig:Classes}
\end{figure}

The entanglement of pure three-qubit states is a fascinating topic.
The system of three qubits is the smallest nontrivial multipartite system,
and, thanks to some mathematical coincidences,
important exotic results could be obtained here.
The first was the \emph{Coffman-Kundu-Wootters} (CKW) \emph{inequality}~\cite{Coffman-2000},
\begin{equation}
\label{eq:CKW}
c_{a|b}^2(\psi) + c_{a|c}^2(\psi) \leq c_{a|bc}^2(\psi),
\end{equation}
expressing the \emph{monogamy of entanglement}, that is,
the entanglement of a qubit with another restricts its entanglement with the third one.
Here $c_{a|b}^2(\psi)$ is the \emph{con\-cur\-rence-squared}~\cite{Wootters-1998},
measuring bipartite entanglement inside the two-qubit subsystem $ab$;
and $c_{a|bc}^2(\psi)$ is the \emph{con\-cur\-rence-squared}~\cite{Coffman-2000},
measuring bipartite entanglement of the whole three-qubit system with respect to the $a|bc$ split.
The difference of the two sides of the inequality
defines the famous \emph{three-tangle} $\tau$~\cite{Coffman-2000},
\begin{equation}
\label{eq:CKWtau}
c_{a|b}^2(\psi) + c_{a|c}^2(\psi) + \tau(\psi) = c_{a|bc}^2(\psi).
\end{equation}
This measures~\cite{Dur-2000} the \emph{residual entanglement},
which is the tripartite entanglement that cannot be accounted for by the entanglement inside the two-qubit subsystems~\cite{Coffman-2000}.
All of these quantities are based on \emph{homogeneous polynomials} of the coefficients of the state vector,
we will recall these precisely later.
The other celebrated result for three-qubit systems was that
there are two \emph{equivalence classes} of genuine three-qubit entanglement
with respect to \emph{stochastic local operations and classical communications} (SLOCC)~\cite{Dur-2000},
called W and GHZ, represented by the W and GHZ states
$\ket{\psi_\text{W}}=\bigl(\ket{100}+\ket{010}+\ket{001}\bigr)/\sqrt{3}$ and
$\ket{\psi_\text{GHZ}}=\bigl(\ket{000}+\ket{111}\bigr)/\sqrt{2}$.
In this way the whole SLOCC classification was carried out for three qubit pure states (Figure~\ref{fig:Classes}).
The operational importance of SLOCC-equivalence is that equivalent states can be used in the same LOCC protocols, although with usually different probability of success.
Two states are SLOCC-equivalent 
 if and only if they can be transformed into each other by local invertible transformations~\cite{Dur-2000}.
The three-tangle $\tau$ plays an important role here, separating these two equivalence classes,
as $\tau(\psi)\neq0$ exactly for state vectors $\ket{\psi}$ contained in the GHZ class~\cite{Dur-2000}.

Almost a decade after these fundamental results came the 
discovery of the so called \emph{FTS approach of three-qubit entanglement}~\cite{Borsten-2009},
coming from the point of view of the black hole/qubit correspondence~\cite{Borsten-2009b,Levay-2011b,Borsten-2012},
with direct relations to entanglement theory~\cite{Levay-2010b,Levay-2011a}.
The three-qubit Hilbert space is isomorphic to 
the \emph{Freudenthal triple system} (FTS) over the cubic Jordan algebra $\field{C}\oplus\field{C}\oplus\field{C}$;
and it turned out that the structure of the three-qubit SLOCC classification (Figure~\ref{fig:Classes})
is naturally captured by the structure of the FTS,
as state vectors of different SLOCC classes correspond to elements of different \emph{ranks} in the FTS~\cite{Borsten-2009}.
This also means an ordering of the SLOCC classes, as triseparable, biseparable, W and GHZ classes
correspond to FTS-ranks $1$, $2$, $3$, and $4$.
This correspondence is based on that the \emph{automorphism group} of the FTS, leaving the FTS-rank invariant, is $\LieGrp{SL}(2,\field{C})^{\times3}$, 
the same group which describes SLOCC-equivalence (up to normalization).
Later the operational significance of the ordering of the SLOCC classes was also discovered~\cite{Borsten-2013}.
Namely, the optimal success rates of winning a three-player nonlocal game~\cite{Watrous-2006},
coming from Mermin's take on the GHZ experiment~\cite{Mermin-1990},
are strictly ordered by the FTS-rank of the shared three-qubit entanglement.
More precisely, the success rate of winning the game 
by the best possible strategy using shared biseparable class state is strictly smaller than
by the best possible strategy using shared W class state, which is strictly smaller than
by the best possible strategy using shared GHZ class state. The last one is $1$, utilizing the $\ket{\psi_\text{GHZ}}$ state, it always wins.
This insight was made possible by the use of the \emph{tensors} by which the FTS rank can be determined~\cite{Borsten-2009,Borsten-2013},
making the FTS approach more than just a different way of the description of the three-qubit SLOCC classification.

Summarizing the situation, 
there are the \emph{three types of entangled SLOCC equivalence classes of three-qubit systems}, the three biseparable ones, the W and the GHZ (Figure~\ref{fig:Classes}),
containing stronger and stronger entanglement in the operational terms of the GHZ game.
Now we propose that
there exist \emph{three types of entanglement of three-qubit systems}, bipartite (entanglement with respect to bipartite split), W and GHZ.
It is important to realize that
the existence of the \emph{three types of entanglement} and
the existence of the \emph{three types of entangled SLOCC classes}
are two different issues.
Any given state contains some of the three types of entanglement (Figure~\ref{fig:Classes}):
bipartite entanglement is contained in all the biseparable, W and GHZ classes;
W-entanglement is contained in the W and GHZ classes;
GHZ-entanglement is contained only in the GHZ class.
We support our claim by constructing a faithful polynomial measure of W-entanglement $\omega$;
alongside the three-tangle $\tau$, which is a faithful polynomial measure of GHZ-entanglement;
and the bipartite concurrence $c_{a|bc}$, which is a faithful polynomial measure of bipartite entanglement.
The significance of our W measure $\omega$ is that $\tau$, $\omega$ and $c_{a|bc}$ together are \emph{ordered}:
the bipartite entanglement content $c_{a|bc}$ of a state is always larger than the W content $\omega$, which is always larger than the GHZ content $\tau$.
Then GHZ entanglement can be considered stronger than W entanglement, which is stronger than bipartite entanglement,
in accordance with the \emph{principle} that \emph{the measure of a stronger type of entanglement should be lower for any given state}.
This principle is the same as \emph{multipartite monotonicity} in the partial separability classification of mixed states~\cite{Szalay-2015b,Szalay-2017,Szalay-2019}.

\section{Preliminaries}
\label{sec:3qb}

We label the three qubits with $1,2,3$,
and we also use the \emph{variables} $a,b,c\in\set{1,2,3}$.
A formula or sentence containing these label variables are always understood for all \emph{different} $a,b,c\in\set{1,2,3}$
without writing this out explicitly.
Let us have the vector in the three-qubit Hilbert space written in the computational basis as
$\ket{\psi}=\sum_{ijk=0}^1\psi^{ijk}\ket{ijk}\in\mathcal{H}_1\otimes\mathcal{H}_2\otimes\mathcal{H}_3$,
and its norm $n(\psi):=\norm{\psi}=\sqrt{\sum_{ijk=0}^1\abs{\psi^{ijk}}^2}$.
We also use the operators
$\rho:=\proj{\psi}$,
$\rho_a:=\tr_{bc}(\rho)$,
$\rho_{bc}:=\tr_a(\rho)$.
If $\ket{\psi}$ is \emph{normalized}, $n(\psi)=1$, then it is a \emph{state vector}, and the operators are \emph{states}, or \emph{density operators};
but on some occasions it is more convenient to release this constraint.
We also have the \emph{local unitary}, \emph{local special linear} and \emph{local general linear} groups 
$\LieGrp{LU}:=\LieGrp{U}(\mathcal{H}_1)\times\LieGrp{U}(\mathcal{H}_2)\times\LieGrp{U}(\mathcal{H}_3)$,
$\LieGrp{LSL}:=\LieGrp{SL}(\mathcal{H}_1)\times\LieGrp{SL}(\mathcal{H}_2)\times\LieGrp{SL}(\mathcal{H}_3)$ and 
$\LieGrp{LGL}:=\LieGrp{GL}(\mathcal{H}_1)\times\LieGrp{GL}(\mathcal{H}_2)\times\LieGrp{GL}(\mathcal{H}_3)$, respectively.
Two quantum states are SLOCC-equivalent if they can be converted into each other by the use of SLOCC.
It turns out that for pure states, SLOCC-equivalence coincides with LGL-equivalence~\cite{Dur-2000}.
The SLOCC classes are then the orbits of the $\LieGrp{LGL}$ group.
For three-qubit pure states these are~\cite{Dur-2000}
the $\text{Null}$ class of the \emph{zero vector};
the ${1|2|3}$ class of \emph{fully separable or triseparable vectors},
which are of the form $\ket{\psi_1}\otimes\ket{\psi_2}\otimes\ket{\psi_3}$ with $\ket{\psi_a}\in\mathcal{H}_a$;
the ${a|bc}$ class of \emph{biseparable vectors},
which are of the form $\ket{\psi_a}\otimes\ket{\psi_{bc}}$ with $\ket{\psi_a}\in\mathcal{H}_a$ and $\ket{\psi_{bc}}\in\mathcal{H}_b\otimes\mathcal{H}_c$ with $\ket{\psi_{bc}}\neq\ket{\psi_b}\otimes\ket{\psi_c}$; and
the W and GHZ classes of \emph{tripartite entangled vectors}, which cannot be written in a product form.
(See Figure~\ref{fig:Classes} for illustration.)
The representing elements of the classes are
\begin{subequations}
\label{eq:SLOCCrepr}
\begin{align}
    \ket{\psi_\text{Null}} &:= 0,\\
    \ket{\psi_{1|2|3}}     &:= \ket{000},\\
    \ket{\psi_{1|23}}      &:= \bigl(\ket{000}+\ket{011}\bigr)/\sqrt{2},\\
    \ket{\psi_{2|13}}      &:= \bigl(\ket{000}+\ket{101}\bigr)/\sqrt{2},\\
    \ket{\psi_{3|12}}      &:= \bigl(\ket{000}+\ket{110}\bigr)/\sqrt{2},\\
    \label{eq:SLOCCrepr.W}
    \ket{\psi_\text{W}}    &:= \bigl(\ket{100}+\ket{010}+\ket{001}\bigr)/\sqrt{3},\\
    \label{eq:SLOCCrepr.GHZ}
    \ket{\psi_\text{GHZ}}  &:= \bigl(\ket{000}+\ket{111}\bigr)/\sqrt{2}.
\end{align}
\end{subequations}
$\LieGrp{LGL}$-equivalence is $\LieGrp{LSL}$-equivalence up to a constant factor,
so it is enough to use the $\LieGrp{LSL}$ group when quantum states are considered, and the vectors are renormalized after the transformation anyways.
Note that entanglement measures are invariant under the action of the $\LieGrp{LU}$ group;
and the vanishing of entanglement measures identifying SLOCC classes (being zero on some classes and nonzero on the others) is invariant under the action of the $\LieGrp{LGL}$ group.

Let us have the antisymmetric element in the two copies of the dual of a qubit Hilbert space $\mathcal{H}_a$~\cite{Szalay-2013,Bengtsson-2017}
\begin{equation}
\label{eq:epsilon}
\bra{\epsilon_a} := \sum_{ii'=0}^1 \epsilon_{ii'}\bra{ii'} := \bra{01} - \bra{10}
\in\mathcal{H}_a^*\otimes\mathcal{H}_a^*.
\end{equation}
There are two ways of using this.
Taking $\bra{\epsilon_a}$ as a \emph{linear map} $\bra{\epsilon_a}:\mathcal{H}_a\to\mathcal{H}_a^*$,
it represents the \emph{spin-flip} or time reversal operation~\cite{Wootters-1998}
on the qubit state vectors $\ket{\phi} = \sum \phi^i\ket{i}\in\mathcal{H}_a$ as
$\bra{\tilde{\phi}}:=\epsilon_a\ket{\phi}:=\bra{\epsilon_a}(\Id\otimes\ket{\phi}) = \sum \epsilon_{ii'}\phi^{i'}\bra{i}$,
by which the spin-flipped vector is $\ket{\tilde{\phi}}= \sum \epsilon_{ii'}\overline{\phi^{i'}}\ket{i}$.%
\footnote{Summation is understood for indices occurring twice.}
More generally,
$\bra{\epsilon_a}$ as a \emph{bilinear form} $\bra{\epsilon_a}:\mathcal{H}_a\otimes\mathcal{H}_a\to\field{C}$
is an $\LieGrp{SL}(\mathcal{H}_a)$ invariant one, as for all $A\in\Lin(\mathcal{H}_a)$ we have
\begin{subequations}
\label{eq:epsilontraf}
\begin{equation}
    \bra{\epsilon_a} A\otimes A = \det(A)\bra{\epsilon_a},
\end{equation}
or, written out by indices, $A=\sum A\indexud{i}{j}\ket{i}\bra{j}$ and
\begin{equation}
    \sum\epsilon_{ii'} A\indexud{i}{j}A\indexud{i'}{j'} = (A\indexud{0}{0}A\indexud{1}{1}-A\indexud{1}{0}A\indexud{0}{1})\epsilon_{jj'}.
\end{equation}
\end{subequations}
This means that $\LieGrp{SL}(\mathcal{H}_a)$ tensors~\cite{Borsten-2009} (also scalars in particular)
can be constructed
by acting with $\bra{\epsilon_a}$-s on the pairs of the same $a$ subsystem of the $m$ copies 
$\ket{\psi}^{\otimes m}\in(\mathcal{H}_1\otimes\mathcal{H}_2\otimes\mathcal{H}_3)^{\otimes m}$
of $\ket{\psi}$.

Before turning to three-qubit systems, we may recall the case of two qubits for warmup,
where there is only one parameter characterizing entanglement.
For two-qubit state vectors $\ket{\phi}=\sum_{ij=0}^1\phi^{ij}\ket{ij}\in\mathcal{H}_1\otimes\mathcal{H}_2$,
let us have the \emph{quadratic form} $q(\phi)\in\field{C}$, 
\begin{equation}
\begin{split}
    q(\phi) &:= 2\det(\phi) =(\bra{\epsilon_1}\otimes\bra{\epsilon_2})(\ket{\phi}\otimes\ket{\phi})\\
    &\phantom{:}=\innerprod{\tilde{\phi}}{\phi}=\sum\epsilon_{ii'}\epsilon_{jj'}\phi^{ij}\phi^{i'j'}.
\end{split}
\end{equation}
Here $\bra{\epsilon_a}\in\mathcal{H}_a^*\otimes\mathcal{H}_a^*$ acts on subsystems $a\in\set{1,2}$ of the two copies.
By local operations  $A_a\in\Lin(\mathcal{H}_a)$, it transforms as
\begin{equation}
    q(A_1\otimes A_2\ket{\phi})=\det(A_1)\det(A_2)q(\phi),
\end{equation}
which can be seen from the~\eqref{eq:epsilontraf} transformation of $\bra{\epsilon_a}$,
so it is an $\LieGrp{LSL}$-invariant (scalar) in particular.
It is also invariant to the permutation of the subsystems.
By the quadratic form $q$ we have the \emph{concurrence}~\cite{Wootters-1998}
\begin{equation}
\label{eq:c12}
\begin{split}
    c_{1|2}(\phi) &:= \abs{q(\phi)} = 2\abs{\det(\phi)} = \abs{\innerprod{\tilde{\phi}}{\phi}}\\
    &\phantom{:}= 2\sqrt{\det(\omega_a)} = \sqrt{2\bigl(\tr(\omega_a)^2-\tr(\omega_a^2)\bigr)},
\end{split}
\end{equation}
where $\omega:=\proj{\phi}$ and $\omega_a:=\tr_b(\omega)$,
which can also be interpreted as the magnitude of the overlap with the spin-flip.
It is invariant to $\abs{\det(A_a)}=1$ transformations,
so it is an $\LieGrp{LSL}$-invariant and also an $\LieGrp{LU}$-invariant.
It also inherits the permutation invariance of $q$.
The importance of $c_{1|2}$ is that
$c_{1|2}(\phi)>0$ if and only if $\ket{\phi}$ is entangled,
and it is an entanglement monotone, 
characterizing entanglement in two-qubit pure states.
Among all the two-qubit state vectors, the Bell state $\ket{\phi_\text{B}}:=\bigl(\ket{00}+\ket{11}\bigr)/\sqrt{2}$ maximizes $c_{1|2}$, and $c_{1|2}(\phi_\text{B})=1$.
The same properties hold for the \emph{concurrence-squared} $c_{1|2}^2$, which is the \emph{linear entropy of the reduced state} $\omega_a$.
Note that not this, but the square of the convex roof extension~\cite{Wootters-1998,Coffman-2000} of $c_{1|2}$ appears in the CKW inequality~\eqref{eq:CKW},
measuring entanglement of the generally mixed two-qubit marginals of the pure three-qubit state.

In the three-qubit case, there are more possibilities of using $\bra{\epsilon_a}$,
and we will write only the indexed versions of the quantities.

\begin{table}
\setlength{\tabcolsep}{04pt}
\begin{tabular}{c|c||c|ccc|c|c}
    Class   & rk &   $n$ &    $c_{1|23}$ &    $c_{2|13}$ &    $c_{3|12}$ &  $\omega$ & $\tau$  \\
            &    & $n^4$ & $n^2c_{1|23}$ & $n^2c_{2|13}$ & $n^2c_{3|12}$ & $n\omega$ & $\tau$  \\
    \hline
    \hline
    Null    &  0 & $=0$   & $=0$ & $=0$ & $=0$   & $=0$   & $=0$ \\
    \hline
    $1|2|3$ &  1 & $>0$   & $=0$ & $=0$ & $=0$   & $=0$   & $=0$ \\
    \hline
    $1|23$  &  2 & $>0$   & $=0$ & $>0$ & $>0$   & $=0$   & $=0$ \\
    $2|13$  &  2 & $>0$   & $>0$ & $=0$ & $>0$   & $=0$   & $=0$ \\
    $3|12$  &  2 & $>0$   & $>0$ & $>0$ & $=0$   & $=0$   & $=0$ \\
    \hline
    W       &  3 & $>0$   & $>0$ & $>0$ & $>0$   & $>0$   & $=0$ \\
    \hline
    GHZ     &  4 & $>0$   & $>0$ & $>0$ & $>0$   & $>0$   & $>0$ \\
\end{tabular}
\caption{SLOCC classes and FTS ranks of three-qubit state vectors
identified by the vanishing of the entanglement measures~\eqref{eq:tau},~\eqref{eq:omega} and~\eqref{eq:ca} (first row).
We also have the entanglement ordering~\eqref{eq:entorder.nn},
that is, the degree-$4$ measures (second row) are decreasing in the columns from the left to the right.}
\label{tab:222classes}
\end{table}

For three-qubit state vectors $\ket{\psi}$,
let us have the \emph{quartic form} $q(\psi)\in\field{C}$~\cite{Borsten-2009},
\begin{equation}
\label{eq:q}
\begin{split}
    &q(\psi) := -2\Det(\psi)\\
    &\quad\phantom{:}=\sum \epsilon_{ii'}\epsilon_{jj'}\epsilon_{kk'}\epsilon_{ll'}\epsilon_{mm'}\epsilon_{nn'}
        \psi^{lmn}\psi^{im'n'}\psi^{l'jk}\psi^{i'j'k'}\\
    &\quad\phantom{:}=\sum \epsilon_{ii'}\epsilon_{jj'}\epsilon_{kk'}\epsilon_{ll'}\epsilon_{mm'}\epsilon_{nn'}
        \psi^{lmn}\psi^{l'jn'}\psi^{im'k}\psi^{i'j'k'}\\
    &\quad\phantom{:}=\sum \epsilon_{ii'}\epsilon_{jj'}\epsilon_{kk'}\epsilon_{ll'}\epsilon_{mm'}\epsilon_{nn'}
        \psi^{lmn}\psi^{l'm'k}\psi^{ijn'}\psi^{i'j'k'}.
\end{split}
\end{equation}
Here $\Det$ is Cayley's hyperdeterminant~\cite{Cayley-1845,Gelfand-1994},
and the equivalence of the three forms can be seen by the permutation invariance, which can be seen by the explicit form~\cite{Coffman-2000,Szalay-2026}.
By local operations $A_a\in\Lin(\mathcal{H}_a)$, it transforms as
\begin{equation}
\label{eq:qltraf}
\begin{split}
    &q(A_1\otimes A_2 \otimes A_3\ket{\psi})\\
    &\quad= \det(A_1)^2\det(A_2)^2\det(A_3)^2 q(\psi),
\end{split}
\end{equation}
which can be seen from the~\eqref{eq:epsilontraf} transformation of $\bra{\epsilon_a}$,
so it is an $\LieGrp{LSL}$-invariant (scalar) in particular.
It is also invariant to the permutation of the subsystems,
which can be seen directly from the equivalence of the three forms~\eqref{eq:q}.
By the quadratic form $q(\psi)$ we have the \emph{three-tangle}~\cite{Coffman-2000,Dur-2000}
\begin{equation}
\label{eq:tau}
    \tau(\psi) := 2\abs{q(\psi)} = 4\abs{\Det(\psi)}.
\end{equation}
It is invariant to $\abs{\det(A_a)}=1$ transformations,
so it is an $\LieGrp{LSL}$-invariant and also an $\LieGrp{LU}$-invariant.
It also inherits the permutation invariance of $q(\psi)$.
The importance of $\tau$ is that
$\tau(\psi)>0$ if and only if $\ket{\psi}$ is contained in the GHZ class~\cite{Dur-2000}
(see Table~\ref{tab:222classes} for a summary),
and $\tau$ is an entanglement monotone~\cite{Dur-2000}.

For three-qubit state vectors $\ket{\psi}$,
let us have the \emph{cubic tensor} $\ket{T(\psi)}\in\mathcal{H}_1\otimes\mathcal{H}_2\otimes\mathcal{H}_3$~\cite{Borsten-2009},
\begin{equation}
\label{eq:T}
\begin{split}
    \ket{T(\psi)} &:= \sum T(\psi)^{ijk}\ket{ijk}\\
    &\phantom{:}= \sum -\epsilon_{ll'}\epsilon_{mm'}\epsilon_{nn'}\psi^{imn}\psi^{lm'n'}\psi^{l'jk}\ket{ijk}\\
    &\phantom{:}= \sum -\epsilon_{ll'}\epsilon_{mm'}\epsilon_{nn'}\psi^{ljn}\psi^{l'mn'}\psi^{im'k}\ket{ijk}\\
    &\phantom{:}= \sum -\epsilon_{ll'}\epsilon_{mm'}\epsilon_{nn'}\psi^{lmk}\psi^{l'm'n}\psi^{ijn'}\ket{ijk}.
\end{split}
\end{equation}
Here the equivalence of the three forms comes by construction~\cite{Borsten-2009}, and can also be seen by the explicit form~\cite{Szalay-2026}.
By local operations $A_a\in\Lin(\mathcal{H}_a)$, it transforms as
\begin{equation}
\label{eq:Tltraf}
\begin{split}
    &\ket{T(A_1\otimes A_2 \otimes A_3\ket{\psi})} \\
    &\quad= \det(A_1)\det(A_2)\det(A_3) A_1\otimes A_2\otimes A_3 \ket{T(\psi)},
\end{split}
\end{equation}
which can be seen from the~\eqref{eq:epsilontraf} transformation of $\bra{\epsilon_a}$,
so it is a tensor transforming as $(\mathbf{2},\mathbf{2},\mathbf{2})$ under the $\LieGrp{LSL}$ group%
\footnote{This notation shows the representations of the three copies of $\LieGrp{SL}(2,\field{C})$,
by which the tensors transform.
$\mathbf{1}$ is the $1$-dimensional trivial representation,
$\mathbf{2}$ is the $2$-dimensional defining representation,
$\mathbf{3}$ is the $3$-dimensional representation acting on the symmetric subspace of the tensor product representation of the given copy.}
if $\ket{\psi}$ transforms as $(\mathbf{2},\mathbf{2},\mathbf{2})$.
It is also invariant to the permutation of the subsystems,
which can be seen directly from the equivalence of the three forms~\eqref{eq:T}.
By the cubic tensor $\ket{T(\psi)}$ let us have its norm
\begin{equation}
\label{eq:omega}
    \omega(\psi) := 2\norm{T(\psi)}.
\end{equation}
It is \emph{not} an $\LieGrp{LSL}$-invariant,
however, it is an $\LieGrp{LU}$-invariant,
and its vanishing is still $\LieGrp{LSL}$-invariant.
It also inherits the permutation invariance of $\ket{T(\psi)}$.
The importance of $\omega$ is that
$\omega(\psi)>0$ if and only if $\ket{\psi}$ is three-qubit entangled, that is, contained in the W or GHZ class~\cite{Borsten-2009,Szalay-2012}
(see Table~\ref{tab:222classes} for a summary),
while its entanglement monotonicity is one of our results, see the next section.

For three-qubit state vectors $\ket{\psi}$,
let us also have the symmetric \emph{quadratic tensors} $\ket{\gamma_a(\psi)}\in \mathcal{H}_a\otimes\mathcal{H}_a$~\cite{Borsten-2009},
\begin{subequations}
\label{eq:gammaa}
\begin{align}
    \ket{\gamma_1(\psi)} &:= \sum  \epsilon_{mm'}\epsilon_{nn'} \psi^{imn}\psi^{i'm'n'} \ket{ii'},\\
    \ket{\gamma_2(\psi)} &:= \sum  \epsilon_{ll'}\epsilon_{nn'} \psi^{ljn}\psi^{l'j'n'} \ket{jj'},\\
    \ket{\gamma_3(\psi)} &:= \sum  \epsilon_{ll'}\epsilon_{mm'} \psi^{lmk}\psi^{l'm'k'} \ket{kk'}.
\end{align}
\end{subequations}
The symmetry is because $\epsilon_b\otimes\epsilon_c$ is symmetric as $\epsilon$ is antisymmetric, and can also be seen by the explicit forms~\cite{Szalay-2026}.
By local operations $A_a\in\Lin(\mathcal{H}_a)$, these transform as
\begin{equation}
\label{eq:Galtraf}
\begin{split}
    &\ket{\gamma_a(A_1\otimes A_2 \otimes A_3\ket{\psi})} \\
    &\quad = \det(A_b)\det(A_c) A_a\otimes A_a \ket{\gamma_a(\psi)},
\end{split}
\end{equation}
which can be seen from the~\eqref{eq:epsilontraf} transformation of $\bra{\epsilon_a}$,
so these are tensors transforming as 
$(\mathbf{3},\mathbf{1},\mathbf{1})$, $(\mathbf{1},\mathbf{3},\mathbf{1})$ and $(\mathbf{1},\mathbf{1},\mathbf{3})$  under the $\LieGrp{LSL}$-group 
if $\ket{\psi}$ transforms as $(\mathbf{2},\mathbf{2},\mathbf{2})$.
Permutation of the subsystems is carried by the lower $a$ indices.
By the quadratic tensors $\ket{\gamma_a(\psi)}$ we have the \emph{concurrence} for all the bipartite splits~\cite{Coffman-2000,Borsten-2009,Szalay-2012} 
\begin{equation}
\label{eq:ca}
\begin{split}
    c_{a|bc}(\psi) &:= \sqrt{\norm{\gamma_b(\psi)}^2 + \norm{\gamma_c(\psi)}^2} \\
    &\phantom{:}= \sqrt{2\bigl(\tr(\rho_a)^2-\tr(\rho_a^2)\bigr)} = 2\sqrt{\det(\rho_a)}.
\end{split}
\end{equation}
These are \emph{not} $\LieGrp{LSL}$-invariants,
however, these are $\LieGrp{LU}$-invariants,
and their vanishing is still $\LieGrp{LSL}$-invariant.
These also inherit the permutation covariance of $\ket{\gamma_a(\psi)}$.
The importance of $c_{a|bc}$ is that
$c_{a|bc}(\psi)>0$ if and only if $\ket{\psi}$ is $a|bc$-entangled, that is, contained in the $b|ac$, $c|ab$, W or GHZ class~\cite{Borsten-2009,Szalay-2012}
(see Table~\ref{tab:222classes} for a summary),
and it is an entanglement monotone,
characterizing bipartite entanglement in three-qubit pure states with respect to the $a|bc$ split.
The same properties hold for the \emph{concurrence-squared} $c_{a|bc}^2$, which is the \emph{linear entropy of the reduced state} $\rho_a$.

It is also useful to express the $\LieGrp{LU}$-invariants $\tau$, $\omega$ and $c_{a|bc}$
by the usual set of algebraically independent three-qubit LU-invariants%
\footnote{Note that we shifted the numbering here from $1,\dots,6$ to $0,\dots,5$, 
since it is more expressive to have $I_1$, $I_2$, $I_3$ in~\eqref{eq:invsI.a} labeled directly by the subsystem index.}%
~\cite{Sudbery-2001,Acin-2001}
\begin{subequations}
\label{eq:invsI}
\begin{align}
    I_0(\psi) &:= \tr(\rho) \equiv \tr(\rho_a) \equiv \norm{\psi}^2,\\
    \label{eq:invsI.a}
    I_a(\psi) &:= \tr(\rho_a^2),\\ 
    \label{eq:invsI.4}
    I_4(\psi) &:= 3\tr(\rho_{bc}(\rho_b\otimes\rho_c)) - \tr(\rho_b^3) - \tr(\rho_c^3),\\
    I_5(\psi) &:= \abs{\Det(\psi)}^2.
\end{align}
\end{subequations}
The three $\LieGrp{LU}$-invariants $I_a$ (together with $I_0$ in the nonnormalized case)
carry one-qubit (local) information only.
The invariant $I_4$ is the Kempe invariant \cite{Kempe-1999,Sudbery-2001},
it is the same for all different $\{b,c\}\subset\{1,2,3\}$ labels,
so it carries two-qubit information only (including also one-qubit information)
which is uniform in the system. 
The invariant $I_5$ is the absolute value of Cayley's hyperdeterminant~\cite{Cayley-1845,Gelfand-1994},
it carries three-qubit information (including also one- and two-qubit information).
The invariants $I_4$ and $I_5$ characterize hidden nonlocality,
that is, nonlocal information which cannot be accessed locally,
which is a resource in quantum cryptography~\cite{Kempe-1999}.
The $\LieGrp{LU}$-invariants $c_{a|bc}$, $\omega$ and $\tau$ can be expressed then
by the $\LieGrp{LU}$-invariants $I_{\dots}$ above, as%
\footnote{This can be seen by the careful use of the identity $\epsilon_{ii'}\epsilon_{ll'} = \delta_{il}\delta_{i'l'} - \delta_{il'}\delta_{i'l}$
for indices corresponding to the same subsystem $a$.
This is also the way of showing the equalities in the \eqref{eq:c12} and \eqref{eq:ca} formulas of the concurrences $c_{1|2}$ and $c_{a|bc}$.}
\begin{subequations}
\label{eq:invsbyI}
\begin{align}
    n^2        &= I_0,\\
    c_{a|bc}^2 &= 2\bigl(I_0^2-I_a\bigr),\\
    \omega^2   &= \frac83 I_4 + \frac{10}{3}I_0^3 - 2I_0\bigl(I_1+I_2+I_3\bigr),\\
    \tau^2     &= 4I_5.
\end{align}
\end{subequations}
Then it is apparent that
$n$ and $c_{a|bc}$ carry one-qubit, $\omega$ uniform two-qubit and $\tau$ three-qubit information.

\section{Results and discussion}
\label{sec:Results}

Our main results are as follows. 
The proofs are given in the Appendix.\\[11pt]
(i) The $\LieGrp{LU}$-invariant $\omega$ is an \emph{entanglement monotone},
that is, nonincreasing on average under pure LOCC maps \cite{Vidal-2000,Horodecki-2001}.\\
(ii) The \emph{entanglement ordering} holds, that is,
\begin{subequations}
\label{eq:entorder}
\begin{equation}
\label{eq:entorder.n}
    0 \leq \tau \leq \omega \leq c_{a|bc} \leq 1
\end{equation}
for normalized (state) vectors, or 
\begin{equation}
\label{eq:entorder.nn}
    0 \leq \tau \leq n\omega \leq n^2 c_{a|bc} \leq n^4
\end{equation}
\end{subequations}
in general, where $n(\psi)=\norm{\psi}$.\\
(iii) The following \emph{maximizations} hold.\\ 
Among all the state vectors, 
$\ket{\psi_\text{GHZ}}$ maximizes $\tau$,
and $\tau(\psi_\text{GHZ})=1$.\\
Among all the non-GHZ class state vectors (closure of class W, $\tau(\psi)=0$),
$\ket{\psi_\text{W}}$ maximizes $\omega$,
and $\omega(\psi_\text{W})=4/(3\sqrt3)$.\\
Among all the $a|bc$ and $1|2|3$ class state vectors (closure of class $a|bc$, $\tau(\psi)=\omega(\psi)=c_{a|bc}(\psi)=0$),
$\ket{\psi_{a|bc}}:=\ket{0}_a\otimes\ket{\phi_\text{B}}_{bc}$ maximizes $c_{b|ac}$ and $c_{c|ab}$,
and $c_{b|ac}(\psi_{a|bc})=c_{c|ab}(\psi_{a|bc})=1$. \\

\begin{figure}\centering
\includegraphics{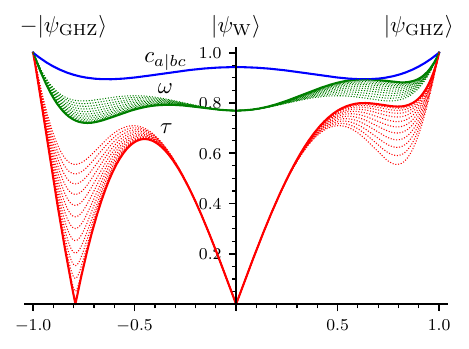}
\caption{Illustration of the entanglement ordering~\eqref{eq:entorder}
for the superposition $\ket{\psi_{x,\phi}} := \sqrt{1-x^2} e^{i\phi}\ket{\psi_\text{W}} + x\ket{\psi_\text{GHZ}}$
for $\phi=0,\dots,\pi/6$ in equidistant steps,
solid line is for $\phi=0$, then dotted lines are in the order of $\phi$.
Some particular points are
$\tau(\psi_{x,\phi})=\omega(\psi_{x,\phi})=c_{a|bc}(\psi_{x,\phi})=1$ at $x=\pm1$ (GHZ state),
$\tau(\psi_{x,\phi})=0<\omega(\psi_{x,\phi})=\sqrt{16/27}<c_{a|bc}(\psi_{x,\phi})=\sqrt{8/9}$ at $x=0$ (W state),
$\omega(\psi_{x,0})=c_{a|bc}(\psi_{x,\phi})=2/\sqrt{5}$ at $x=\sqrt{2/5}$, and 
$\tau(\psi_{x,0})=0$ at $x=0$ and at $x= -\sqrt{4\sqrt[3]{2}/(3+4\sqrt[3]{2})}$.
For $\tau$ see also~\cite{Lohmayer-2006}, with a slightly different parametrization.}
\label{fig:toc1}
\end{figure}

There are several important points here,
let us begin with the measuring of bipartite entanglement,
that is, entanglement with respect to a bipartite split of the whole system.
The theory of measures of pure bipartite entanglement is well understood~\cite{Vidal-2000,Horodecki-2001},
any unitary invariant concave function of the reduced state of a subsystem
is a pure entanglement monotone~\eqref{eq:pureentmon}, measuring the entanglement of that subsystem with the rest of the system.
The concurrence-squared $c_{a|bc}^2$ was given as the \emph{linear entropy} of the reduced state~\eqref{eq:ca},
being the $2(\tr(\rho_a)^2-\tr(\rho_a^2))$ normalized Tsallis entropy of parameter $q=2$, which is unitary invariant and concave.
The point here is that
in our scheme we have the concurrence $c_{a|bc}$ instead of the concurrence-squared $c_{a|bc}^2$
for the role of measuring bipartite entanglement.
It is easy to see that an \emph{increasing concave} function of an entanglement monotone is an entanglement monotone~\eqref{eq:pureentmon},
so not only $c_{a|bc}^2$ but also $c_{a|bc}=\sqrt{c_{a|bc}^2}$ is an entanglement monotone.

So we have the entanglement monotones $c_{a|bc}$, $\omega$ and $\tau$.
Coming from homogeneous polynomials,

these are positively homogeneous functions%
\footnote{A function $f:\mathcal{H}\to\field{C}$ is \emph{positively homogeneous} of degree $k\in\field{R}$ if $f(c\psi)=c^kf(\psi)$ for all $0<c\in\field{R}$,
while it is \emph{homogeneous} of degree $k\in\field{Z}$ if the same holds for all $0\neq c\in\field{C}$.}
of degree $2$, $3$ and $4$,
and the entanglement ordering~\eqref{eq:entorder.nn} is of degree $4$.%
\footnote{Sometimes we call these loosely ``polynomial measures'' for short,
when the point is to distinguish from the von Neumann/Rényi entropy or relative entropy based ``entropic'' measures, containing logarithm.
Note that Tsallis entropy based measures fit into the first category.}
Homogeneous entanglement monotone functions,
being nonincreasing on average~\eqref{eq:pureentmon},
are expected not to be of arbitrary high degree
(see Theorem I.~in \cite{Eltschka-2012}, concerning a special class of functions).
We used that an increasing concave function of an entanglement monotone is an entanglement monotone,
however, this is not the case for convex functions.
Here we have that neither $\tau^2$ nor $\omega^2$ nor $c_{a|bc}^4$ is an entanglement monotone,%
\footnote{This can be seen by constructing counterexamples.}
so, although the entanglement ordering~\eqref{eq:entorder.nn} is equivalent to $0\leq \tau^2\leq n^2\omega^2\leq n^4c_{a|bc}^2\leq n^8$
of degree-$8$ functions,
only $c_{a|bc}^2$ is an entanglement measure in this.
The point here is that
both the CKW equality~\eqref{eq:CKWtau} and the entanglement ordering~\eqref{eq:entorder.nn} are written for degree-$4$ functions,
however, the former relates $\tau$ with $c_{a|bc}^2$,
while the latter with $n^2c_{a|bc}$,
so these represent different kinds of relations among the quantities.
On the other hand, we have $\tau \leq c_{a|bc}^2$ from the CKW equality~\eqref{eq:CKWtau},
which is stronger than $\tau \leq n^2 c_{a|bc}$ from the entanglement ordering~\eqref{eq:entorder.nn},
however, $n\omega \not\leq c_{a|bc}^2$,%
\footnote{This can be seen by constructing counterexamples.}
so $c_{a|bc}^2$ cannot be used instead of $n^2c_{a|bc}$ in the entanglement ordering~\eqref{eq:entorder.nn}.

About the invariance properties, we have that $c_{a|bc}$, $\omega$ and $\tau$ are $\LieGrp{LU}$-invariants, but only $\tau$ is an $\LieGrp{LSL}$-invariant.
Note that $\LieGrp{LSL}$-invariant measures vanish on W-class states, since these vanish for the $\LieGrp{LSL}$-normal form $\ket{\psi_\text{W}}$ \cite{Verstraete-2003}.
The measures $\omega$ and $c_{a|bc}$ are \emph{not} $\LieGrp{LSL}$-invariants, $\delta$-contractions%
\footnote{$\delta$-contractions are the $\LieGrp{U(2)}$-invariant index contractions as in $\sum \delta_{ii'}\psi^{ijk}\overline{\psi^{i'mn}}$ with complex conjugation,
similarly to the $\LieGrp{SL(2)}$-invariant $\epsilon$-contractions as in $\sum \epsilon_{ii'}\psi^{ijk}\psi^{i'mn}$ \cite{Szalay-2026}.}
in the norms~\eqref{eq:omega}, \eqref{eq:ca} spoil that property.
Note also that no measure of $a|bc$-entanglement could be $\LieGrp{LSL}$-invariant,
the $\LieGrp{LSL}$-invariance of the two-qubit concurrence $c_{1|2}$ was an exception, not the rule.
This is because the measures of bipartite entanglement in pure states are exactly the unitary invariant concave functions of the reduced state~\cite{Vidal-2000,Horodecki-2001},
and the state reduction means $\delta$-contractions, spoiling $\LieGrp{LSL}$-invariance.
On the other hand, let us emphasize that, although $\LieGrp{LSL}$-invariance is not a defining property of entanglement measures,
the $\LieGrp{LSL}$ structure itself, that is, applying $\epsilon$-contractions,
is still rather important in our construction of entanglement measures.
It is an interesting question whether this is general, or there are different, still significant approaches independent of the $\LieGrp{LSL}$ structure.
The $\LieGrp{LU}$-invariance is, on the other hand, mandatory~\cite{Vidal-2000,Horodecki-2001}.

Note also that the $\epsilon$-contractions and $\delta$-contractions can directly be translated to the language of quantum circuits,
making possible the experimental measurement of $c_{a|bc}$, $\omega$ and $\tau$ on quantum computers~\cite{Szalay-2026}.
The probability of a particular outcome of the measurement of the qubit registers of the implementations
gives the value of $c_{a|bc}$, $\omega$ and $\tau$ directly.

After these technical points, let us turn to the different multipartite entanglement properties.

First let us emphasize again that \emph{bipartite entanglement} is
entanglement with respect to a bipartite split of the whole three-qubit system,
not to be confused with \emph{two-qubit entanglement} inside two-qubit subsystems.
The latter can be zero also for a zero measure subset of the GHZ class,
containing for example the $\ket{\psi_\text{GHZ}}$ state~\eqref{eq:SLOCCrepr.GHZ}, having separable two-qubit subsystems.
We have \emph{three kinds of bipartite entanglement}, with respect to the three splits $1|23$, $2|13$ and $3|12$.
On the other hand, we have also \emph{three kinds of biseparable classes}, $1|23$, $2|13$ and $3|12$,
containing only two-qubit entanglement.
Note, however, that bipartite entanglement is contained in all the entangled classes.
For example, $1|23$-entanglement, that is, bipartite entanglement with respect to the split $1|23$ 
is contained in the biseparable classes $2|13$ and $3|12$, in the W and in the GHZ class,
but \emph{not} in the class $1|23$, there are $2|13$- and $3|12$-entanglement instead.
This might be slightly confusing,
as the same symbols label the biseparable classes and the entanglement with respect to bipartite splits,
Figure~\ref{fig:Classes} and the relevant column of Table~\ref{tab:222classes} helps to see the pattern.

\begin{figure}\centering
\includegraphics{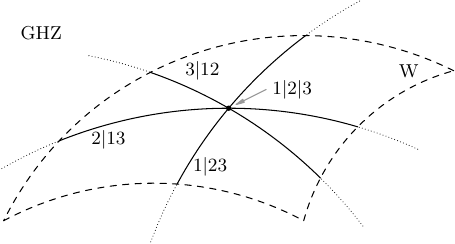}
\caption{Illustration of the geometry of the SLOCC classification of three-qubit state vectors.
Classes lower in the hierarchy are of zero measure in the closure of classes higher in the hierarchy.
Class $1|2|3$ is represented by a dot in the middle;
the closures of classes $a|bc$ are represented by three curves;
the closure of class $\text{W}$ is represented by the two-dimensional surface with dashed borders;
the closure of class $\text{GHZ}$, which is the whole state space, is represented by the three-dimensional ``space'' of the figure.}
\label{fig:Strat}
\end{figure}

The SLOCC classes are disjoint subsets of the space of state vectors 
$\mathcal{S}^{15}\subset\mathcal{H}_1\otimes\mathcal{H}_2\otimes\mathcal{H}_3$, the $15$-dimensional sphere given by $n(\psi)=1$.
The class $1|2|3$ is of zero measure in the closures of the classes $a|bc$,
which are of zero measures in the closure of class W,
which is of zero measure in the closure of class GHZ, which is the whole state space.
These are algebraic varieties given by the zero-loci of the corresponding invariants (Table~\ref{tab:222classes}),
and the classification itself is a stratification~\cite{Bengtsson-2017} of the space of the state vectors (Figure~\ref{fig:Strat}).
This is what establishes the robustness of the ordering $a|bc$, W and GHZ of the SLOCC classes:
a stronger class should not be contained in the closure of a weaker.%
\footnote{See also the corresponding reasoning in the SLOCC-based classification of~\emph{mixed} states~\cite{Acin-2001b}.
The same holds naturally in the partial separability classification~\cite{Szalay-2015b,Szalay-2017,Szalay-2019}, when it is restricted to pure states:
a weaker class is contained as a zero measure subset in the closure of a stronger.}
Note that the GHZ game~\cite{Borsten-2013} does not prove, only motivates and illustrates this ordering.
First of all, one may possibly be able to formulate another three-qubit game, where, e.g., a given W-class state $\ket{\psi_*}$ is more valuable than all the GHZ-class states.
Then, however, one could find GHZ-class states with success rate arbitrarily close to the success rate of $\ket{\psi_*}$.
The point is that even if such a game could be formulated, it would not be useful in identifying the ordering of the classes.
The importance of the GHZ game~\cite{Borsten-2013} is that it points at a concrete operational situation where the ordering is apparent.

\begin{figure}\centering
\includegraphics{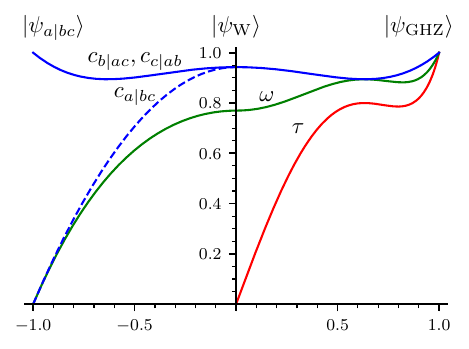}
\caption{Illustration of the entanglement ordering~\eqref{eq:entorder} and the maximizations of the measures
for the superpositions $\ket{\psi_{x\geq0}} := \sqrt{1-x^2}\ket{\psi_\text{W}} + x\ket{\psi_\text{GHZ}}$ for coefficients $x\geq0$
and $\ket{\psi_{x\leq0}} := \sqrt{1-x^2}\ket{\psi_\text{W}} - x\ket{\psi_{a|bc}'}$ for coefficients $x\leq0$,
where we use $\ket{\psi_{a|bc}'} := \ket{1}_a\otimes\bigl(\ket{01}+\ket{10}\bigr)_{bc}/\sqrt{2}$, $\LieGrp{LU}$-equivalent to $\ket{\psi_{a|bc}}$,
in order that the latter superposition stay in the closure of class W.}
\label{fig:toc2}
\end{figure}

Having the \emph{entanglement measures} $c_{a|bc}$, $\omega$ and $\tau$ in hand, 
it is meaningful to consider \emph{three types of three-qubit entanglement}, bipartite, W and GHZ,
in parallel to the \emph{three types of SLOCC classes of entangled three-qubit states}.
More precisely, let W-entanglement and GHZ-entanglement themselves be defined to be what are measured by $\omega$ and $\tau$,
in the same manner as bipartite entanglement is defined in this case to be what is measured by $c_{a|bc}$.
Having also the \emph{entanglement ordering}~\eqref{eq:entorder} in hand,
it is meaningful to establish the order of the \emph{three types of three-qubit entanglement}: bipartite is weaker than W, which is weaker than GHZ.
This is in parallel to the order of the \emph{three types of SLOCC classes of entangled three-qubit states}, 
where the same order was motivated by the possible usefulness in the GHZ game~\cite{Borsten-2013},
and was argued by the geometry of the classes (Figure~\ref{fig:Strat}).
States of class $a|bc$ contain only bipartite entanglement (not $a|bc$ but $b|ac$ and $c|ab$ entanglement); 
states of class W contain all the three kinds of bipartite, and also \emph{a smaller amount} of W-entanglement;
states of class GHZ contain all the three kinds of bipartite, \emph{a smaller amount} of W-, and also \emph{an even smaller amount} of GHZ-entanglement.
See Figure~\ref{fig:Classes} and Table~\ref{tab:222classes} for the pattern and Figures~\ref{fig:toc1} and~\ref{fig:toc2} for numerical illustration.

The \emph{entanglement ordering}~\eqref{eq:entorder} expresses that GHZ-entanglement is stronger than W, which is stronger than bipartite,
in the sense that any state may contain a smaller amount of the stronger form of entanglement than of the weaker.
This is the same principle for three-qubit pure states
as \emph{multipartite monotonicity} in the \emph{partial separability classification} of multipartite mixed states~\cite{Szalay-2015b,Szalay-2017,Szalay-2019},
where the ``strength'' of partial entanglement properties (labeled by partition ideals) are partially ordered by the so called \emph{refinement} order,
and the relative entropies of those properties were smaller for the stronger form of entanglement.
Note that in the case of multipartite monotonicity~\cite{Szalay-2015b,Szalay-2017,Szalay-2019},
the strength and ordering of entanglement properties are well established, and the ordering of the measures follows.
In the case of entanglement ordering~\eqref{eq:entorder},
we have the ordering of the measures, and we would like to infer to the strength and ordering of the types of entanglement.
To this, it is obviously needed that the types of entanglement are measured by the measures $c_{a|bc}$, $\omega$ and $\tau$, which are ordered~\eqref{eq:entorder}.
Of course one could deform these measures, or construct entirely new measures, which satisfy the vanishing conditions (Table~\ref{tab:222classes}), but spoil the ordering~\eqref{eq:entorder}.
We can support the use of $c_{a|bc}$, $\omega$ and $\tau$ by that these measures are rather natural, being just norms of the $\LieGrp{LSL}$-tensors, naturally characterizing the classes themselves.
Moreover, the stronger type of entanglement is measured by a measure of higher degree, what also seems remarkable.
The situation of this reasoning resembles the situation in entanglement monogamy~\eqref{eq:CKW}, which holds for the measures $c^2_{a|c}$ and $c^2_{a|bc}$.
Of course one could deform these measures, or construct entirely new measures, which satisfy the vanishing conditions, but spoil the monogamy.
However, the concurrence-squared $c^2$ is rather natural to measure bipartite entanglement, and we consider entanglement monogamy itself as an established feature of quantum theory.

Note also that the structure of 
the entanglement measures $c_{a|bc}$, $\omega$ and $\tau$ in the ordering of entanglement of three-qubit pure states~\eqref{eq:entorder},
and the measures natural in the partial separability classification of mixed states~\cite{Szalay-2015b,Szalay-2017,Szalay-2019}
are different.
From the point of view of the partial separability classification,
$\omega$ could also be considered as a measure of genuine multipartite entanglement (GME),
since it is nonzero exactly for nonseparable state vectors, that is, W- and GHZ-class vectors (Table~\ref{tab:222classes}).
In the SLOCC-extended~\cite{Acin-2001b,Szalay-2012} three-qubit partial separability classification~\cite{Szalay-2015b,Szalay-2017,Szalay-2019},
we would measure how much a state is \emph{not} fully separable, \emph{not} $a|bc$-separable, \emph{not} biseparable, and \emph{not} non-GHZ-entangled
by entropic measures~\cite{Szalay-2015b,Szalay-2017}.
Or we would measure GME by forming the minimum~\cite{Ma-2011,Szalay-2015b} or geometric mean or other means~\cite{Szalay-2015b} of bipartite concurrences or other bipartite measures.
Our present point of view is different,
here we measure how much a state is $a|bc$-, W- and GHZ-entangled \emph{directly},
applying polynomial measures $c_{a|bc}$, $\omega$ and $\tau$.
These do not come from a general construction, but arise naturally, as being deeply related to the SL structure $\epsilon$.

From the \emph{maximizations} we have that,
in the whole state space,
the $\LieGrp{LU}$-orbit of $\ket{\psi_\text{GHZ}}$ maximizes all the three types of entanglement,
\begin{subequations}
\label{eq:max}
\begin{equation}
\label{eq:max.GHZ}
    \tau(\psi_\text{GHZ})=\omega(\psi_\text{GHZ})=c_{a|bc}(\psi_\text{GHZ})=1;
\end{equation}
in the closure of class W (given by $\tau(\psi)=0$),
the $\LieGrp{LU}$-orbit of $\ket{\psi_\text{W}}$ maximizes W-entanglement, but not $a|bc$-entanglement,
\begin{equation}
\label{eq:max.W}
    \omega(\psi_\text{W})=\sqrt{16/27}<c_{a|bc}(\psi_\text{W})=\sqrt{8/9}<1;
\end{equation}
and in the closure of class $a|bc$ (given by $\tau(\psi)=\omega(\psi)=c_{a|bc}(\psi)=0$), 
the $\LieGrp{LU}$-orbit of $\ket{\psi_{a|bc}}$ maximizes $b|ac$ and $c|ab$-entanglement,
\begin{equation}
\label{eq:max.avbc}
    c_{b|ac}(\psi_{a|bc})=c_{c|ab}(\psi_{a|bc})=1.
\end{equation}
\end{subequations}
See Figure~\ref{fig:toc2} for illustration.
So, in the closure of class W, the maximal value of $\omega$ is less than $1$, and this forbids the maximization of $c_{b|ac}$ for any particular split
($c_{b|ac}$ can reach $1$ for $\ket{\psi_{a|bc}}$ in the closure of class W),
contrary to the whole state space, where the maximizations of $\tau$, $\omega$ and $c_{a|bc}$ occur simultaneously~\eqref{eq:max.GHZ}.
On the other hand, it can also be seen that in the closure of class W,
the three concurrences $c_{a|bc}$ cannot reach $1$ \emph{simultaneously},
the simultaneous maximal value is $\sqrt{8/9}$, taken for the W state $\ket{\psi_\text{W}}$;
and the \emph{average concurrence} $(c_{1|23}+c_{2|13}+c_{3|12})/3$ is also maximized then.
Also, if in the closure of class W, the concurrence $c_{b|ac}$ for \emph{a particular} split takes $1$, then the maximal value of $\omega$ is $1/\sqrt{2}$.
(The proof of these are also given in the Appendix.)
These suggest that there are further nontrivial bounds among these entanglement measures beyond
the CKW inequality~\eqref{eq:CKW} and the CKW equality~\eqref{eq:CKWtau},
  relating entanglement inside two-qubit subsystems with entanglement of the whole system; and
the entanglement ordering~\eqref{eq:entorder},
  relating the three types of entanglement of the whole three-qubit system.

\section{Summary and remarks}
\label{sec:Summary}

We proposed that
there exist \emph{three types of three-qubit entanglement}, bipartite, W and GHZ,
which are also ordered, bipartite is weaker than W, which is weaker than GHZ.
These are given by the measures $c_{a|bc}$, $\omega$ and $\tau$, respectively,
which are 
faithful (in the sense of Table~\ref{tab:222classes}) ordered~\eqref{eq:entorder} polynomial measures,
natural from the point of view of the FTS approach.
It is important to realize that
the existence of the \emph{three types of entanglement} and 
the existence of the \emph{three types of entangled SLOCC equivalence classes}
are two different issues.
Any given entangled state contains some (maybe all) of the three types of entanglement,
accordingly to its SLOCC class, following the pattern illustrated in Figure~\ref{fig:Classes} and given in Table~\ref{tab:222classes}.

The most important open question is the operational meaning of the measures $c_{a|bc}$, $\omega$ and $\tau$
as resource quantifiers in some nonlocal protocols, tasks or games.
Note that the GHZ game~\cite{Borsten-2013} is only about the best possible strategies,
the usefulness of any given state is not characterized.
We only have that there exist W class states which are more useful than all the biseparable class states,
and there exist GHZ class states which are more useful than all the W class states in the GHZ game.
However, some GHZ class state can be less useful than some W class states,
and some W class state can be less useful than some biseparable class states.

We would also like to argue that the possible lack of operational meaning does not undermine the importance of the measures.
Note that the three-tangle $\tau$ has no known operational meaning either, while being important beyond doubt.
Neither the concurrence-squared $c_{a|bc}^2$ nor the concurrence $c_{a|bc}$ has any known \emph{direct} operational meaning;
they are just increasing functions of the von Neumann entropy of the reduced state.
This latter is the entanglement entropy, having the direct operational meaning of the asymptotic LOCC convertibility rate into Bell-pairs~\cite{Bennett-1996}.
Nevertheless, even if no direct operational meaning of $c_{a|bc}^2$ and $\tau$ can be found,
they are important for a different reason, namely, the CKW equality~\eqref{eq:CKWtau} about the distribution of entanglement could be formulated by them.
The von Neumann entropy of the reduced state, or any other function of the concurrence would not work here~\cite{Coffman-2000}.
Similarly, even if no direct operational meaning of $c_{a|bc}$, $\omega$ and $\tau$ can be found,
they are important for a different reason, namely, the entanglement ordering~\eqref{eq:entorder} could be formulated by them.

As a closing remark,
we note that the construction presented here is rather specific for three qubits,
and does not seem to have a straightforward generalization for arbitrary qu$d$its, or for arbitrary number of subsystems.
In this manner we call this exotic.
The FTS approach turned out to be fruitful also for some further exotic multipartite quantum systems~\cite{Levay-2008,Vrana-2009,Sarosi-2014a,Sarosi-2014b,Sarosi-2014c,Holweck-2016}.
We note that $\omega^2$ appears also in the twistor-geometric approach of three-qubit entanglement~\cite{Levay-2005},
the invariant $\omega_{ABC}$ there is $-\omega^2$ above,
it is nonpositive and it is not monotone on average under pure LOCC.

\section*{Author contributions}
All the one author contributed equally to this work.
Moreover, the author actually contributed even more equally than himself.
This is why the author is listed alphabetically.
No AI was used in producing the work.
The author verifies that he is a human. 

\begin{acknowledgments}
We thank Péter Lévay and Frédéric Holweck for discussions,
and two anonymous reviewers for their valuable questions and suggestions.
Financial support 
of the Hungarian National Research, Development and Innovation Office
within the grant K-134983,
within the `Frontline' Research Excellence Programme KKP-133827,
and within the Quantum Information National Laboratory of Hungary is gratefully acknowledged.
We happily acknowledge the support of the wonderful Bach performances of Marta Czech and Sir András Schiff.
\end{acknowledgments}

\appendix
\section{Proofs}
\label{sec:Proofs}

\subsection{Entanglement monotonicity of $\omega$}
\label{sec:Proofs.i}

Here we consider normalized vectors, $\norm{\psi}^2=1$.

First, recall that for state vectors $\ket{\psi}\in\mathcal{H}=\bigotimes_a \mathcal{H}_a$,
the function $f:\mathcal{H}\to\field{R}$ is an \emph{entanglement monotone}~\cite{Vidal-2000,Horodecki-2001}
by definition if it is nonincreasing on average under pure LOCC~\cite{Horodecki-2001},
that is, LOCC channel where the suboperations are of Kraus-rank 1.
Formally,
\begin{equation}
\label{eq:pureentmon}
    \sum_{\mu = 1}^m p_\mu f(\psi_\mu')\leq f(\psi),
\end{equation}
where the output state vectors $\ket{\psi_\mu'} = \frac1{\sqrt{p_\mu}} M_\mu\ket{\psi}$
with outcome probabilities $p_\mu = \norm{M_\mu\ket{\psi}}^2\neq0$
are given by the product Kraus operators
$M_\mu=\bigotimes_a M_{\mu,a}$
such that $\sum_{\mu = 1}^m M_{\mu,a}^\dagger M_{\mu,a}=\Id_a$.
We have some simplifications here~\cite{Dur-2000},
as any such protocol can be generated by two-outcome protocols, $m=2$,
where the operation on the system is done in only one subsystem at a time,
$M_\mu=M_{\mu,a}\otimes \bigotimes_{b\neq a}\Id_b$,
so it is enough to consider this case in~\eqref{eq:pureentmon} for all $a$ in general.

Turning to the function $\omega: \mathcal{H}\to\field{R}$,
being permutation invariant~\eqref{eq:omega},~\eqref{eq:T}, 
it is enough to consider operations acting only on a fixed qubit,
$M_\mu=A_\mu\otimes\Id\otimes\Id$.
Let us have the singular value decomposition of $A_\mu$ as
$A_\mu=U_\mu\sqrt{D_\mu}V^\dagger\in\Lin(\mathcal{H}_1)$ with the unitaries
$U_\mu,V\in\mathrm{U}(\mathcal{H}_1)$,
and the diagonal positive operator 
\begin{subequations}
\begin{equation}
    D_\mu = a_\mu\proj{0}+b_\mu\proj{1} \in \Lin(\mathcal{H}_1)
\end{equation}
with
\begin{equation}
\label{eq:ab}
    0\leq a_\mu, b_\mu,\quad a_1+a_2= b_1+b_2=1.
\end{equation}
\end{subequations}
(Note that $V$ is not $\mu$-dependent, which holds for the two-outcome case,
as we have $A_1^\dagger A_1+A_2^\dagger A_2 = \Id$,
from which $V_2^\dagger V_1D_1V_1^\dagger V_2 = \Id-D_2 = D_1$,
leading to $V_2^\dagger V_1=\Id$, that is, $V_1=V_2$.)

First, let us have the vector
\begin{subequations}
\label{eq:xi}
\begin{equation}
\label{eq:xi.vector}
    \ket{\xi} := \ket{0}\otimes\ket{\xi_0}+\ket{1}\otimes\ket{\xi_1}
    := V^\dagger\otimes\Id\otimes\Id\ket{\psi},
\end{equation}
which is normalized, and
\begin{equation}
\label{eq:xi.norm}
    x_i:=\norm{\xi_i}^2,\quad x_0+x_1=\norm{\xi}^2=1.
\end{equation}
\end{subequations}
With these we have the outcome probabilities
\begin{equation}
\label{eq:pmu}
\begin{split}
    p_\mu &= \bra{\psi}A_\mu^\dagger A_\mu\otimes\Id\otimes\Id\ket{\psi}\\
    &= \bra{\psi}V D_\mu V^\dagger\otimes\Id\otimes\Id\ket{\psi}\\
    &= \bra{\xi} D_\mu \otimes\Id\otimes\Id\ket{\xi}
    = x_0 a_\mu + x_1 b_\mu.
\end{split}
\end{equation}

Second, let us have the vector
\begin{subequations}
\label{eq:upsilon}
\begin{equation}
\label{eq:upsilon.vector}
    \ket{\zeta} := \ket{0}\otimes\ket{\zeta_0}+\ket{1}\otimes\ket{\zeta_1}
    := \frac{V^\dagger\otimes\Id\otimes\Id\ket{T(\psi)}}{\omega(\psi)/2},
\end{equation}
which is normalized by definition~\eqref{eq:omega}, and
\begin{equation}
\label{eq:upsilon.norm}
    y_i:=\norm{\zeta_i}^2,\quad y_0+y_1=\norm{\zeta}^2=1.
\end{equation}
\end{subequations}
Note that we can assume $\omega(\psi)\neq0$,
since if $\omega(\psi)=0$ then $\omega(\psi'_\mu)=0$ as well by~\eqref{eq:Tltraf} and~\eqref{eq:omega},
then~\eqref{eq:pureentmon} holds.
With these we have for the output state vectors
\begin{equation}
\label{eq:omegapsimup}
\begin{split}
    \omega(\psi_\mu')
    &\equalsref{eq:omega} 2\bignorm{T\big(A_\mu\otimes\Id\otimes\Id\ket{\psi}/\sqrt{p_\mu}\big)} \\ 
    &\equalsref{eq:Tltraf} 2 \bignorm{\det(A_\mu)A_\mu\otimes\Id\otimes\Id\ket{T(\psi)}/\sqrt{p_\mu}^3}\\
    &\equals \frac2{\sqrt{p_\mu}^3} \abs{\det(A_\mu)} \bignorm{A_\mu\otimes\Id\otimes\Id\ket{T(\psi)}}\\
    &\equals \frac2{\sqrt{p_\mu}^3} \det\bigl(\sqrt{D_\mu}\bigr) \bignorm{\sqrt{D_\mu} V^\dagger\otimes\Id\otimes\Id\ket{T(\psi)}}\\
    &\equalsref{eq:upsilon.vector} \frac1{\sqrt{p_\mu}^3} \det\bigl(\sqrt{D_\mu}\bigr) \bignorm{\sqrt{D_\mu} \otimes\Id\otimes\Id\ket{\zeta}\omega(\psi)}\\
    &\equals \frac1{\sqrt{p_\mu}^3} \sqrt{a_\mu b_\mu \bra{\zeta} D_\mu\otimes\Id\otimes\Id\ket{\zeta}}\omega(\psi) \\ 
    &\equalsref{eq:upsilon.norm} \frac1{\sqrt{p_\mu}^3} \sqrt{a_\mu b_\mu (y_0 a_\mu + y_1 b_\mu) }\omega(\psi).
\end{split}
\end{equation}

Having the value \eqref{eq:omegapsimup} for the output state vectors
and the outcome probabilities \eqref{eq:pmu},
the entanglement monotonicity~\eqref{eq:pureentmon} for $\omega$ takes the form
\begin{equation}
\label{eq:claim1}
    \sum_{\mu=1}^2\sqrt{a_\mu b_\mu\frac{y_0 a_\mu + y_1 b_\mu}{x_0 a_\mu + x_1 b_\mu}} \leq 1.
\end{equation}
As $a_\mu x_0 + b_\mu x_1$ is the convex combination of $a_\mu$ and $b_\mu$ by \eqref{eq:xi.norm},
we have
\begin{equation*}
\begin{split}
    &\sqrt{a_\mu b_\mu\frac{y_0 a_\mu + y_1 b_\mu}{x_0 a_\mu + x_1 b_\mu}}\\
    &\quad\leq x_0 \sqrt{a_\mu b_\mu\frac{y_0 a_\mu + y_1 b_\mu}{a_\mu}}
       + x_1 \sqrt{a_\mu b_\mu\frac{y_0 a_\mu + y_1 b_\mu}{b_\mu}}\\
    &\quad= \bigl(x_0\sqrt{b_\mu}+x_1\sqrt{a_\mu}\bigr)\sqrt{y_0 a_\mu + y_1 b_\mu}
\end{split}
\end{equation*}
by the convexity of $u\mapsto 1/\sqrt{u}$.
Then~\eqref{eq:claim1} takes the form
\begin{equation*}
\begin{split}
    &\sum_{\mu=1}^2\sqrt{a_\mu b_\mu\frac{y_0 a_\mu + y_1 b_\mu}{x_0 a_\mu + x_1 b_\mu}}\\
    &\quad\leq x_0\bigl(\sqrt{b_1}\sqrt{y_0 a_1 + y_1 b_1} + \sqrt{b_2}\sqrt{y_0 a_2 + y_1 b_2}\bigr) \\
    &\quad+    x_1\bigl(\sqrt{a_1}\sqrt{y_0 a_1 + y_1 b_1} + \sqrt{a_2}\sqrt{y_0 a_2 + y_1 b_2}\bigr) \\
    &\quad\leq x_0 + x_1 = 1,
\end{split}
\end{equation*}
where the second inequality is due to the Cauchy-Bunyakovsky-Schwarz (CBS) inequality in $\field{R}^2$,
for example, the first parenthesis is
\begin{equation*}
\begin{split}
&\bigl(\sqrt{b_1},\sqrt{b_2}\bigr)\cdot\bigl(\sqrt{y_0 a_1 + y_1 b_1},\sqrt{y_0 a_2 + y_1 b_2}\bigr)\\
&\quad\leq\sqrt{b_1+b_2}\sqrt{y_0 a_1 + y_1 b_1+y_0 a_2 + y_1 b_2}\\
&\quad= \sqrt{b_1+b_2}\sqrt{y_0(a_1+a_2) + y_1(b_1+b_2)}=1,
\end{split}
\end{equation*}
by~\eqref{eq:ab} and~\eqref{eq:upsilon.norm}.

\subsection{Entanglement ordering}
\label{sec:Proofs.ii}

Here we consider arbitrary vectors $\ket{\psi}$, not necessarily normalized.

To see the bound 
\begin{equation}
\label{eq:claim.ii1}
    \tau^2\leq n^2\omega^2
\end{equation}
in the entanglement ordering~\eqref{eq:entorder.nn},
we can read off from \eqref{eq:q} and \eqref{eq:T}
that the quadratic form can be written as the inner product
\begin{equation}
    q(\psi)= -\innerprod{\tilde{\psi}}{T(\psi)}
\end{equation}
between $\ket{T(\psi)}$ and 
the spin-flipped state vector $\ket{\tilde{\psi}}=(\epsilon_1\otimes\epsilon_2\otimes\epsilon_3)\sum_{ijk}\overline{\psi^{ijk}}\ket{ijk}$.
Then by the CBS inequality we have
\begin{equation}
    \abs{q(\psi)}^2 = \abs{\innerprod{\tilde{\psi}}{T(\psi)}}^2 \leq \norm{\psi}^2 \norm{T(\psi)}^2, 
\end{equation}
as $\norm{\tilde{\psi}}^2=\norm{\psi}^2$,
so the definitions~\eqref{eq:tau} and~\eqref{eq:omega} lead to~\eqref{eq:claim.ii1}.
Note that the necessary and sufficient condition of the equality in~\eqref{eq:claim.ii1} is $\ket{T(\psi)}=z\ket{\tilde{\psi}}$ for a $z\in\field{C}$.

To see the bound 
\begin{equation}
\label{eq:claim.ii2}
    \omega^2\leq n^2 c_{a|bc}^2
\end{equation}
in the entanglement ordering~\eqref{eq:entorder.nn},
we make use of the \eqref{eq:invsbyI} expressions of $\omega^2$ and $c_{a|bc}^2$
in terms of the $\LieGrp{LU}$-invariants~\eqref{eq:invsI}.
Then we have for the difference of the sides of~\eqref{eq:claim.ii2}
\begin{equation*}
\begin{split}
    & n^2 c_{a|bc}^2 - \omega^2 
     \equalsref{eq:invsbyI} 2I_0\bigl(I_b+I_c\bigr) - \frac43 I_0^3 - \frac83 I_4  \\
    &\quad\equalsref{eq:invsI}  2\tr(\rho) \bigl(\tr(\rho_b^2)+\tr(\rho_c^2)\bigr)  - \frac43 \tr(\rho)^3 \\
    &\quad\phantom{\equals} +\frac83\bigl(\tr(\rho_b^3)+\tr(\rho_c^3)\bigr) -8\tr(\rho_{bc}(\rho_b\otimes\rho_c))\\
    &\quad\equals 4\bigl(\tr(\rho_b^3)+\tr(\rho_c^3)\bigr)-8\tr(\rho_{bc}(\rho_b\otimes\rho_c))\\
    &\quad\equals 4\tr\bigl(\rho_{bc}(\rho_b\otimes \Id_c - \Id_b \otimes \rho_c)^2\bigr)\geq0,
\end{split}
\end{equation*}
where we used that $\tr(\rho)=\tr(\rho_b)=\tr(\rho_c)$ obviously, and the identity
$2\tr(A)\tr(A^2)-\frac23\tr(A)^3 = \frac43\tr(A^3)$.
(This can be seen by the Cayley-Hamilton theorem $A^2-\tr(A)A+\det(A)\Id=0$ for dimension $2$,
from which
we have $\tr(A^2)-\tr(A)^2+2\det(A)=0$
by taking the trace, 
and
$\tr(A^3)-\tr(A)\tr(A^2)+\det(A)\tr(A)=0$
by multiplying with $A$ and then taking the trace.)
In the last term 
the inner parenthesis, being the difference of commuting positive operators, is self-adjoint,
so its square is positive,
so we have the trace of the product of two positive operators,
which is nonnegative.
Note that a sufficient condition of the equality in~\eqref{eq:claim.ii2} is $\rho_b=\Id/2$ and $\rho_c=\Id/2$,
and the necessary and sufficient condition is the orthogonality of $\rho_{bc}$ and $(\rho_b\otimes \Id_c - \Id_b \otimes \rho_c)^2$,
which is not particularly expressive.

The bound 
\begin{equation}
\label{eq:claim.ii3}
    c_{a|bc}^2 \leq n^4
\end{equation}
in the entanglement ordering~\eqref{eq:entorder.nn} is somewhat well known,
we just show it for the sake of completeness.
There are many ways to show it,
for instance, it follows from the properties of the Tsallis entropies~\cite{Szalay-2013},
however, we give a direct proof here by the \eqref{eq:invsbyI} expressions of $c_{a|bc}^2$ and $n^2$
in terms of the $\LieGrp{LU}$-invariants~\eqref{eq:invsI}.
Then we have for the difference of the sides of~\eqref{eq:claim.ii3}
\begin{equation*}
    n^4 - c_{a|bc}^2
    \equalsref{eq:invsbyI} 2I_a - I_0^2
    \equalsref{eq:invsI}   2\tr(\rho_a^2) - \tr(\rho_a)^2 \geq 0,
\end{equation*}
where the inequality can be seen by writing out the left-hand side by the $\lambda_1$, $\lambda_2$ eigenvalues of $\rho_a$,
leading to $2(\lambda_1^2+\lambda_2^2)-(\lambda_1+\lambda_2)^2 = (\lambda_1-\lambda_2)^2$,
which is nonnegative.
Note that the necessary and sufficient condition of the equality in~\eqref{eq:claim.ii3} is $\rho_a=\Id/2$.

\subsection{Maximizations}
\label{sec:Proofs.iii}

Here we consider normalized vectors, $\norm{\psi}^2=1$.

To see that 
$\tau$ is maximized by $\ket{\psi_\text{GHZ}}$ among all the state vectors,
we just have that $\tau(\psi_\text{GHZ})=1$,
and this is the maximal value for state vectors by~\eqref{eq:entorder.n}.

To see that 
$\omega$ is maximized by $\ket{\psi_\text{W}}$ among all the non-GHZ class state vectors
is not that straightforward.
Since we are interested in the maximization of an $\LieGrp{LU}$-invariant function over states of $\LieGrp{LU}$-invariant properties (SLOCC-classes),
it is enough to consider the $\LieGrp{LU}$-canonical form of three-qubit vectors~\cite{Acin-2000,Acin-2001}
\begin{subequations}
\label{eq:canonical}
\begin{equation}
\label{eq:canonical.psi}
\begin{split}
    \ket{\psi_\text{C}} := \sqrt{\eta_0}\ket{000}
    + e^{i\theta}\sqrt{\eta_1}\ket{100} &+ \sqrt{\eta_2}\ket{101}\\
               + \sqrt{\eta_3}\ket{110} &+ \sqrt{\eta_4}\ket{111},
\end{split}
\end{equation}
where $0\leq\eta_i\in\field{R}$, $0\leq\theta\leq\pi$, and
\begin{equation}
\label{eq:canonical.norm}
    \norm{\psi_\text{C}}^2 = \eta_0+\eta_1+\eta_2+\eta_3+\eta_4 = 1
\end{equation}
\end{subequations}
is the normalization.
The squared entanglement measures~\eqref{eq:invsbyI} evaluated for this are
\begin{subequations}
\label{eq:invsbyI.canonical}
\begin{align}
    c_{1|23}^2(\psi_\text{C}) &= 4\eta_0(\eta_2+\eta_3+\eta_4), \\
    c_{2|13}^2(\psi_\text{C}) &= 4\eta_0(\eta_3+\eta_4) + 4\abs{\delta}^2, \\
    c_{3|12}^2(\psi_\text{C}) &= 4\eta_0(\eta_2+\eta_4) + 4\abs{\delta}^2, \\
    \omega^2(\psi_\text{C})   &= 4\eta_0\eta_4 \norm{\psi_\text{C}}^2 - 16\eta_0\sqrt{\eta_2\eta_3}\Re(\delta), \\
    \tau^2(\psi_\text{C})     &= 16\eta_0^2\eta_4^2,
\end{align}
\end{subequations}
with $\delta:=\sqrt{\eta_1\eta_4}e^{i\theta} -\sqrt{\eta_2\eta_3}$. 
Non-GHZ state vectors are those for which $\tau(\psi_\text{C})=4\eta_0\eta_4=0$,
which holds if and only if $\eta_0=0$ or $\eta_4=0$.
The case $\eta_0=0$ leads to $\omega(\psi_\text{C})=0$, so we consider the case $\eta_4=0$,
then $\delta=-\sqrt{\eta_2\eta_3}$ and
$\omega(\psi_\text{C})=4\sqrt{\eta_0\eta_2\eta_3}$,
the maximum of which is searched for.
For this, we have
\begin{equation*}
\sqrt[3]{\eta_0\eta_2\eta_3}\leq\frac13(\eta_0+\eta_2+\eta_3)\leq\frac13(\eta_0+\eta_1+\eta_2+\eta_3) = \frac13,
\end{equation*}
where the first inequality is the inequality between the geometric and arithmetic means,
and the equality is the normalization~\eqref{eq:canonical.norm} for the $\eta_4=0$ case.
Both inequalities can be saturated, the first one for $\eta_0=\eta_2=\eta_3$, the second one for $\eta_1=0$.
This gives $\eta_0=\eta_2=\eta_3=1/3$ because of the normalization,
so the maximum is $\omega(\psi_\text{C*})=4/\sqrt{3^3}$,
taken for the state $\ket{\psi_\text{C*}}=\bigl(\ket{000}+\ket{101}+\ket{110}\bigr)/\sqrt{3}$, being $\LieGrp{LU}$-equivalent to $\ket{\psi_\text{W}}$
by the Pauli-X unitary $X=\ketbra{0}{1}+\ketbra{1}{0}$ acting on the first qubit.

We may have some further findings in the closure of class W.
The case $\eta_4=0$ leads to that
$c_{1|23}^2(\psi_\text{C}) = 4\eta_0(\eta_2+\eta_3)$,
$c_{2|13}^2(\psi_\text{C}) = 4\eta_3(\eta_0+\eta_2)$,
$c_{3|12}^2(\psi_\text{C}) = 4\eta_2(\eta_0+\eta_3)$,
taking the same value if and only if $\eta_0=\eta_2=\eta_3\leq1/3$, 
for which the maximum is $8/9$ when $\eta_1=0$.
The average concurrence
$2\bigl(\sqrt{\eta_0(\eta_2+\eta_3)}+\sqrt{\eta_3(\eta_0+\eta_2)}+\sqrt{\eta_2(\eta_0+\eta_3)}\bigr)/3$ is maximized when $\eta_1=0$,
then it is $2(\sqrt{\eta_0(1-\eta_0)}+\sqrt{\eta_2(1-\eta_2)}+\sqrt{\eta_3(1-\eta_3)})/3$, then the CBS inequality
leads to $\eta_0=\eta_2=\eta_3=1/3$.
Also, if only $c_{1|23}^2(\psi_\text{C})=1$, then $\eta_0=1/(4(\eta_2+\eta_3))$ and $\omega^2(\psi_\text{C})=4\eta_2\eta_3/(\eta_2+\eta_3)$,
which takes its maximum $1/2$ for $\eta_2=\eta_3=1/4$, $\eta_0=1/2$, when $\eta_0+\eta_2+\eta_3\leq1$.

To see that 
$c_{b|ac}$ and $c_{c|ab}$ are maximized by $\ket{\psi_{a|bc}}:=\ket{0}_a\otimes\ket{\phi_\text{B}}_{bc}$ among all the $a|bc$ and $1|2|3$ class state vectors,
we just have that $c_{b|ac}(\psi_{a|bc})=c_{c|ab}(\psi_{a|bc})=1$,
and this is the maximal value for state vectors by~\eqref{eq:entorder.n}.

\bibliographystyle{quantum}
\bibliography{3qbW}{}

\end{document}